\documentclass[preprint,12pt]{elsart}

 \usepackage{epsfig}

\usepackage{amssymb}
\usepackage{rotating}
\usepackage{tikz}
\usepackage{lscape}
\usepackage{rotating}
\begin{document}
\begin{frontmatter}
\title{Different seniority states of $^{119-126}$Sn isotopes: shell model description}
\author{}
\author{Praveen C. Srivastava$^{a,}$$^{*}$ and  Bharti Bhoy$^{a}$}
\address{$^{a}$Department of Physics, Indian Institute of Technology Roorkee, Roorkee - 247667, India }
\author{M.J. Ermamatov$^{b}$}
\address{$^{b}$Institute of Nuclear Physics, Ulughbek, Tashkent 
100214, Uzbekistan}
\address{$^{*}$Corresponding author email-id: pcsrifph@iitr.ac.in}

\date{\today}

\begin{abstract}  
In the present work available experimental data up to high-spin states of $^{119-126}$Sn isotopes with different seniority ($v$), including  $v$ = 4, 5, 6, and 7 have been interpreted with shell model, by performing full-fledged shell model calculations in the 50-82 valence shell composed of  
$1g_{7/2}$, $2d_{5/2}$, $1h_{11/2}$, $3s_{1/2}$, and $2d_{3/2}$ orbitals. The results 
have been compared with the available experimental data.
These states are described in terms of broken neutron pairs occupying the $h_{11/2}$ orbital.
Possible configurations of seniority isomers in these nuclei
are discussed. The breaking of three neutron pairs have been responsible
for generating high-spin states.  
The isomeric states $5^-$, $7^-$, $10^+$ and $15^-$ of even Sn isotopes, and isomeric states $19/2^+$, $23/2^+$, $27/2^-$ and $35/2^+$  of odd
Sn isotopes, are described in terms of different seniority. For even-Sn isotopes, the  isomeric states $5^-$, $7^-$, and $10^+$   are due to seniority $v$ = 2; the isomeric state
$15^-$  is due to seniority $v$ = 4,
and in the case of odd-Sn isotopes, the isomeric states $19/2^+$, $23/2^+$, and $27/2^-$  are due to seniority $v$ = 3, and the isomeric state $35/2^+$ in $^{123}$Sn 
is due to  seniority $v$ = 5. These are maximally-aligned spin, which involve successive pair breakings in the $\nu (h_{11/2})$ orbit.

\end{abstract}


\end{frontmatter}

\newpage
\section{Introduction}
The Sn region is one of the important regions, where many experimental and theoretical studies, such as
identification of different isomeric states in Sn isotopes \cite{astier,Iskra,Iskra1,sandlescu,qi,corragio,togashi,snptep,biswas}, Gamow-Teller decay of 
the doubly magic nucleus $^{100}$Sn \cite{hinke}, measurement of electromagnetic properties of different excited states \cite{ekstrom},
upcoming measurements for definite spin assignments \cite{ruiz},  population of high-spin states \cite{astier}, theoretical calculations
of nuclear g factors \cite{zamick}
and {\it ab  initio} study
of lighter Sn isotopes \cite{Morris} are going on. Recent studies report lowering of the $\nu g_{7/2}$ orbital in comparison to the $\nu d_{5/2}$ for the $^{101}$Sn \cite{Darb}. 
It is possible with direct spin assignments, together with magnetic moment measurements, to probe the wave function of the ground states of the  $^{101-107}$Sn isotopes. 
This may help accurately  determine the ordering of the $\nu d_{5/2}-\nu g_{7/2}$ orbitals.  

 The number of particles which are not in pairs
coupled to angular momentum $J$ = 0 is known as seniority ($v$)~\cite {talmi}. 
There are several text book examples where $g_{9/2}$, $h_{11/2}$ and $i_{13/2}$ orbitals are responsible for generating high-spin seniority states.
The $g_{9/2}$ orbital is responsible for $10^+$ and $12^+$ states in the case of $^{94}$Ru and $^{96}$Pd with configuration $\pi g_{9/2}^4$  \cite{piet}.
The $\nu g_{9/2}^4$ configuration is responsible for $8^+$ state of seniority $v=2$ in $^{72}$Ni and $^{74}$Ni \cite{piet}. The high-spin seniority states are due to $h_{11/2}$
orbital in the Sn region. The role of $i_{13/2}$ orbital in 82-126 major shell region is crucial for the seniority $v=4$ states. 
In the case of even $^{200,202}$Pb isotopes the seniority $v=4$ states are $16^+$ [$\nu i_{13/2}^2 \nu f_{5/2}^2$] and 
$17^-,19^-$ [$\nu i_{13/2}^3 \nu f_{5/2}^1$] \cite{fant}.

In the Sn region, the appearance of isomeric states in $N=82$ isotones and $Z=50$ isotopes are very common, in even isotopes the  $10^+$, 
and in odd isotopes $27/2^-$. The role of the $h_{11/2}$ orbital is crucial for the investigation
of these isomers within shell model. 
There are three  different experimental groups which are involved to identify seniority isomers in the Sn isotopes.
Fotiades  \cite{fotiades} group at LBNL, Astier \cite{astier} group at Legnaro and IRes-Strasbourg and Iskra group \cite{Iskra,Iskra1} at Argonne
 have done different experiments to populate isomeric states in odd and even Sn isotopes using fusion-fission reactions.
The high-spin structure above the $10^+$ isomers in $^{118,120,122,124}$Sn reported by Fotiades et al in Ref. \cite{fotiades}.
More complete level schemes in odd and even Sn isotopes with $A$ = 119-126 are populated in $^{12}$C + $^{238}$U and $^{18}$O + $^{208}$Pb fusion-fission reactions
reported by Astier el al \cite{astier}. 
The aim of this experiment was
to built high-spin states above the long-lived isomeric states lying around 4.5 MeV. 
The excited states above the $v=2$ isomers have been populated in even $^{118-128}$Sn isotopes in the fusion-fission reaction \cite{Iskra}, 
while excited states with seniority  $v=$ 3, 5 and 7 have been investigated in odd $^{119-125}$Sn isotopes \cite{Iskra}.
In these experiments \cite{astier,Iskra,Iskra1} for even $^{120,122,124,126}$Sn isotopes the isomeric states are $10^+$, $5^-$, $7^-$ and $15^-$,
while for odd $^{119,121,123,125}$Sn isotopes they are $27/2^-$, $19/2^+$, and $23/2^+$. We have shown these isomeric states in 
the Figs.1 and 2, for even and odd Sn isotopes, respectively. In these figures we have also shown seniority
of these different isomeric states. 
One of the important application of nuclear shell model is to identify the states involving many identical nucleons in the same orbit, 
thus it is possible to identify the high-spin states in the Sn isotopes with $(\nu h_{11/2})^n$ configuration.

\begin{figure}
\vspace{8.4cm}
\begin{center}
\includegraphics[width=10.0cm]{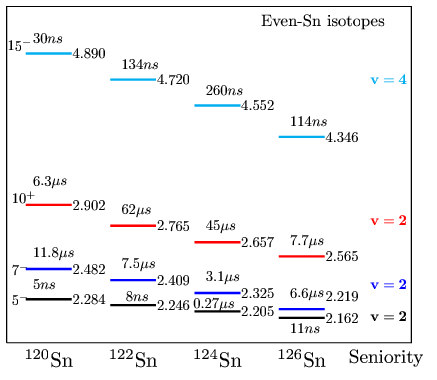}
\end{center}
\vspace{-0.5cm}
\caption{
Different isomeric states \cite{astier,Iskra,Iskra1}  for $^{120,122,124,126}$Sn isotopes.}
\label{Sn_even}
\end{figure}
\begin{figure}
\vspace{8.5cm}
\begin{center}
\includegraphics[width=10.0cm]{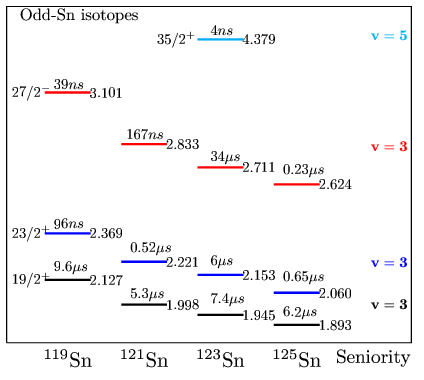}
\end{center}
\vspace{-0.5cm}
\caption{
Different isomeric states  for $^{119,121,123,125}$Sn isotopes~\cite{astier,Iskra,Iskra1}.}
\label{Sn_odd}
\end{figure}

Our focus is to explain different isomeric states in term of seniority as proposed by A. Astier et al. \cite{astier} and 
we have reported additional states in comparison to Refs. \cite{Iskra,Iskra1}. In the Ref. \cite{Iskra,Iskra1},
shell model results for energy  levels are reported only for selected states. Thus, present  work will add more information to earlier shell
model results reported in Refs. \cite{astier,Iskra,Iskra1} and our calculated values for different $B(E2)$ transitions will be very useful for future experiments.

This work is organized as follows:  comprehensive comparison of shell-model
results and experimental data is given in Section 2. 
Configuration of different isomeric states are shown in Section 3.
In Section 4 comparison of the calculated transition probabilities for isomeric states are given. Finally, concluding remarks are drawn in Section 5.

\section{Shell model Hamiltonian and model space}

The shell-model calculations for the Sn isotopes have been performed in the full 50-82 valence shell composed of $1g_{7/2}$, $2d_{5/2}$,
$1h_{11/2}$, $3s_{1/2}$ and $2d_{3/2}$ orbitals without any truncation. 
We have performed calculations with SN100PN interaction due to
Brown {\it et al}~\cite{Mac,PhysRevC.71.044317}.
 The residual two-body interaction was obtained from the CD-Bonn G-matrix within the so-called 
$\hat{Q}$-box folded-diagram theory and further $nn$ part was multiplied  by a factor of 0.90 to improve
the agreement with the experiment for $^{130}$Sn.
The single-particle energies for 
the neutrons are -10.609, -10.289, -8.717, -8.694, and -8.816 MeV for
the $1g_{7/2}$, $2d_{5/2}$, $2d_{3/2}$, $3s_{1/2}$, and $1h_{11/2}$ orbitals, respectively.
The results shown in this work were obtained with the code Antoine \cite{Antoine}. In this region, 
we have previously reported  shell model results for the structural properties 
of some nuclei~\cite{pc1,pc2,pc3,pc4,pc5,pc6,pc7,pc8} using SN100PN interaction.  

\subsection{\label{even} Analysis of spectra of even isotopes of Sn}

Since $^{100}$Sn core is used in this work, neutron excitations are important among the 1$g_{7/2}$, 2$d_{5/2}$, 2$d_{3/2}$, 3$s_{1/2}$ and 1$h_{11/2}$ orbitals for the $^{119-126}$Sn isotopes. The valence neutrons contribute in the
structure of these nuclei because of the $Z$ = 50 shell closure. In this section we perform shell model calculations
for the even-even isotopes in the 50-82 shell, in order to describe the positive and negative parity levels of these nuclei.
The even-even isotopes of Sn are discussed first. The odd isotopes $^{119,121,123,125}$Sn have been studied within shell
model in Ref.~\cite{Iskra}. We  sketch the results for the odd isotopes for the completeness and comparison in subsection~\ref{odd}, including some 
more recently measured states.
 
\begin{figure*}
\begin{center}
\includegraphics[width=15.0cm]{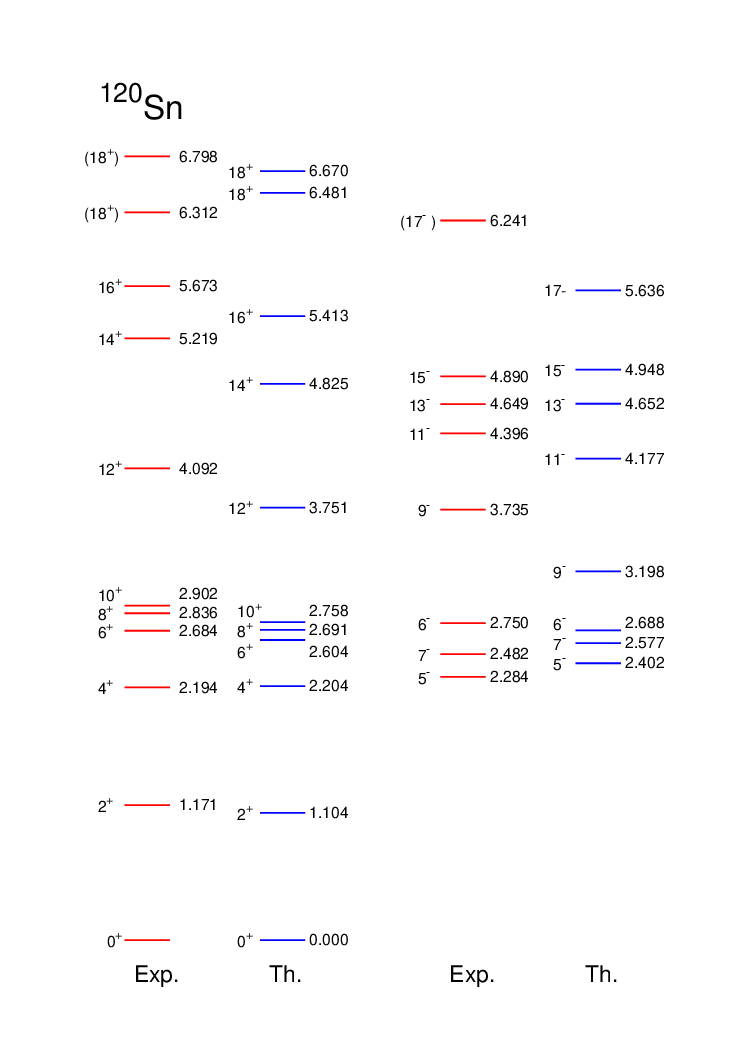}
\end{center}
\caption{
Comparison of experimental \cite{astier,Iskra} and calculated excitation spectra using SN100PN interaction for $^{120}$Sn.}
\label{f_sn120}
\end{figure*}

Comparison of the calculated spectrum of 
$^{120}$Sn with the experimental data is shown in figure~\ref{f_sn120}. 
The calculated 2$^+$ and 4$^+$ levels are 67 keV lower and only 10 keV higher, respectively than those  in the experiment. Then, there are 
gaps both in the experiment and calculation (490 keV and 400 keV, respectively)
 between the 4$^+$ and 6$^+$ levels, calculated one being less. 
 In the calculation, $6^+$, $8^+$ and $10^+$ triple of the levels  is slightly lower and more compressed than the experimental one.
For the negative parity levels, the  $5^-$ and $7^-$ levels are 118 keV and 95 keV higher, respectively, as compared to those of the experimental ones. 
The calculated $9^-$ level is 537 keV
lower than in the experiment. 
In the experiment $11^-$, $13^-$ and $15^-$ levels are almost equidistant, i.e. 253 keV, 241 keV far from each other.


\begin{figure*}
\begin{center}\includegraphics[width=15.0cm]{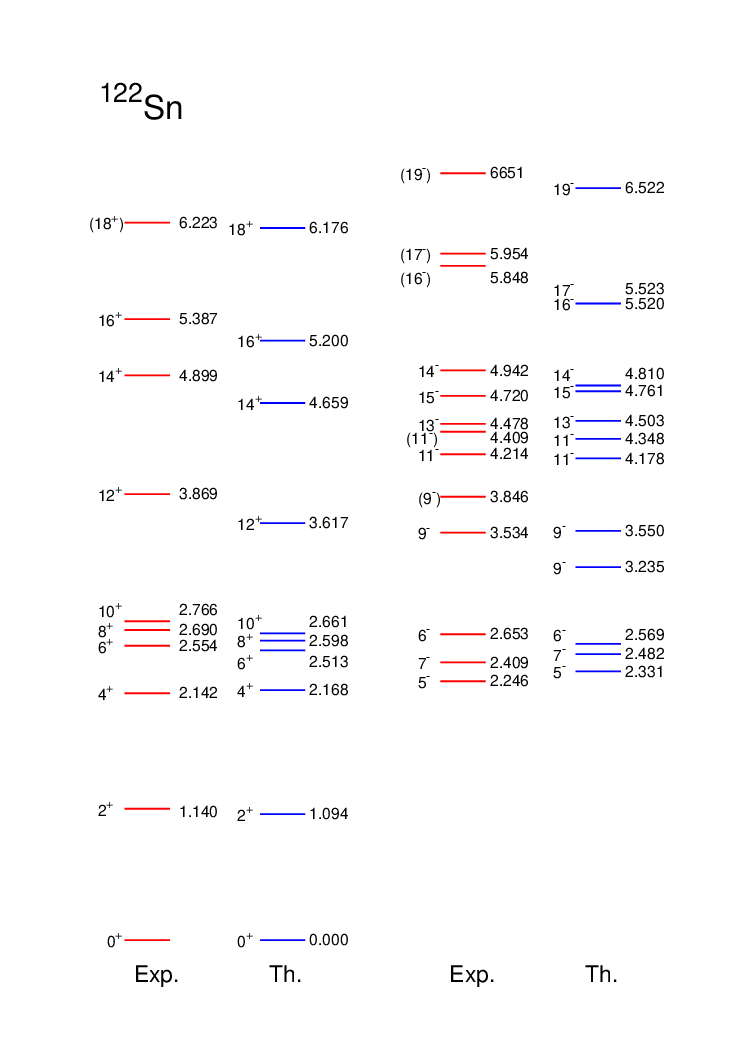}
\end{center} \caption{Comparison of experimental \cite{astier,Iskra} and calculated excitation spectra for $^{122}$Sn using SN100PN interaction.}
\label{f_sn122}
\end{figure*}

Comparison of the calculated values with the experimental data is shown in figure~\ref{f_sn122}. Comparing figures~\ref{f_sn120} and 
\ref{f_sn122} one can see that the positive parity spectrum of the $^{122}$Sn is very similar to that of $^{120}$Sn. 
The 2$^+$ level is predicted 46 keV lower and 4$^+$ level is only 26 keV higher than the experimental values, i.e.
the values of the both energy levels are decreased with respect to the ground state as compared to those of $^{120}$Sn. 
The calculated positive parity levels of $^{122}$Sn are better described by shell model calculation as compared to those of $^{120}$Sn.
For the negative parity levels, the  $5^-$ and $7^-$ levels are 85 and 73 keV higher, respectively, as compared to those of the experimental ones.



\begin{figure*}
\includegraphics[width=15.0cm]{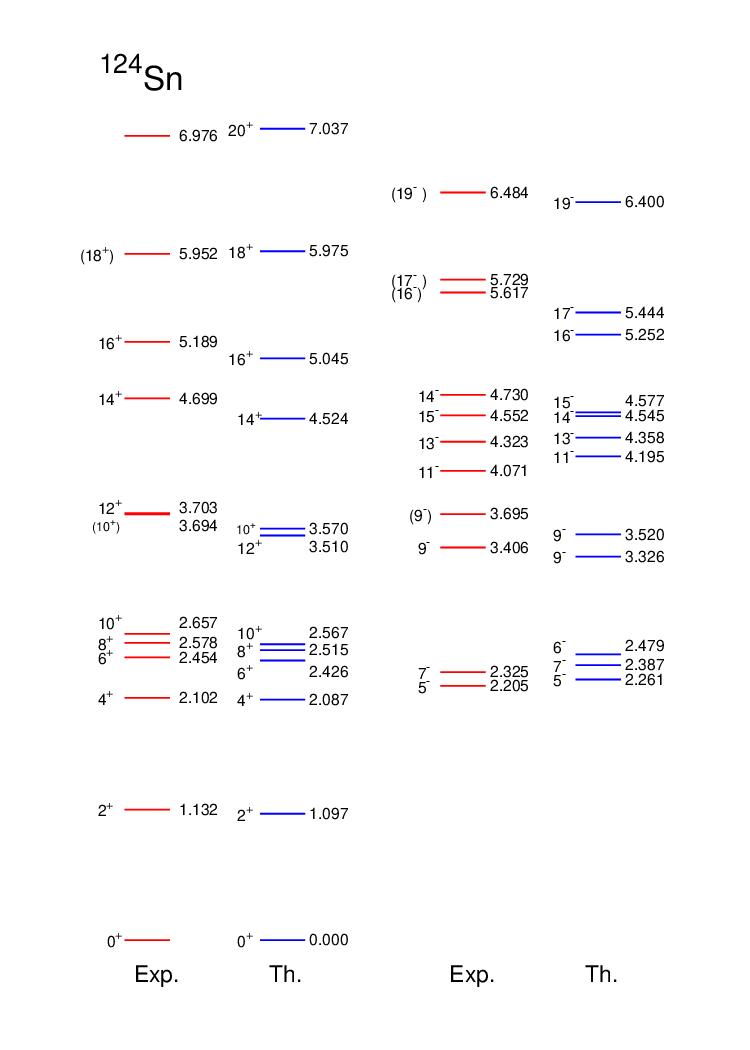}
\caption{
Comparison of experimental \cite{astier,Iskra} and calculated excitation spectra for $^{124}$Sn 
using SN100PN interaction.}
\label{f_sn124}
\end{figure*}
Comparison of the calculated values with the experimental data for the $^{124}$Sn is shown in
figure~\ref{f_sn124}. 
As compared to $^{122}$Sn, in the 
calculation only the energy of $2^+$ level is increased to 3 keV and all other energies of the levels  
are decreased with respect to ground state like in the experiment 
The 2$^+$ and 4$^+$ levels are  only 35 keV and
15 keV lower, respectively,  than the experimental ones which shows better agreement 
as compared to that 
of $^{120,122}$Sn. The values of the respective experimental and calculated energy gaps between the 4$^+$ and 6$^+$ levels are 352 keV and 
339 keV. They
are also in better agreement with the experiment than for $^{120,122}$Sn.
The $6^+$, $8^+$
and $10^+$ triplet of the levels in the calculation is still slightly lower and    
more compressed than in the experiment: 
differences in the values of the experimental 
$6^+$ and $8^+$, and $8^+$ and $10^+$ levels are  124 and 79 keV, respectively, 
while
the calculated values of these differences are 89 and 52 keV, respectively. 
The experimental difference between  the $6^+$ and $8^+$ levels 
is
decreased while, the difference  between the $8^+$ and $10^+$ levels is increased
as compared to that of $^{122}$Sn. Reverse trend is seen in the differences of the 
calculated levels: the difference between  the $6^+$ and $8^+$ level is
increased, the difference between $8^+$ and $10^+$ is decreased as compared to that of 
$^{122}$Sn.

For the negative parity levels, the  $5^-$ and $7^-$ levels 
are 56 and 62 keV higher, respectively,
 as compared to those of the experimental ones. For these two levels the calculations
are clearly better than $^{120,122}$Sn cases.


\begin{figure*}
\includegraphics[width=15.0cm]{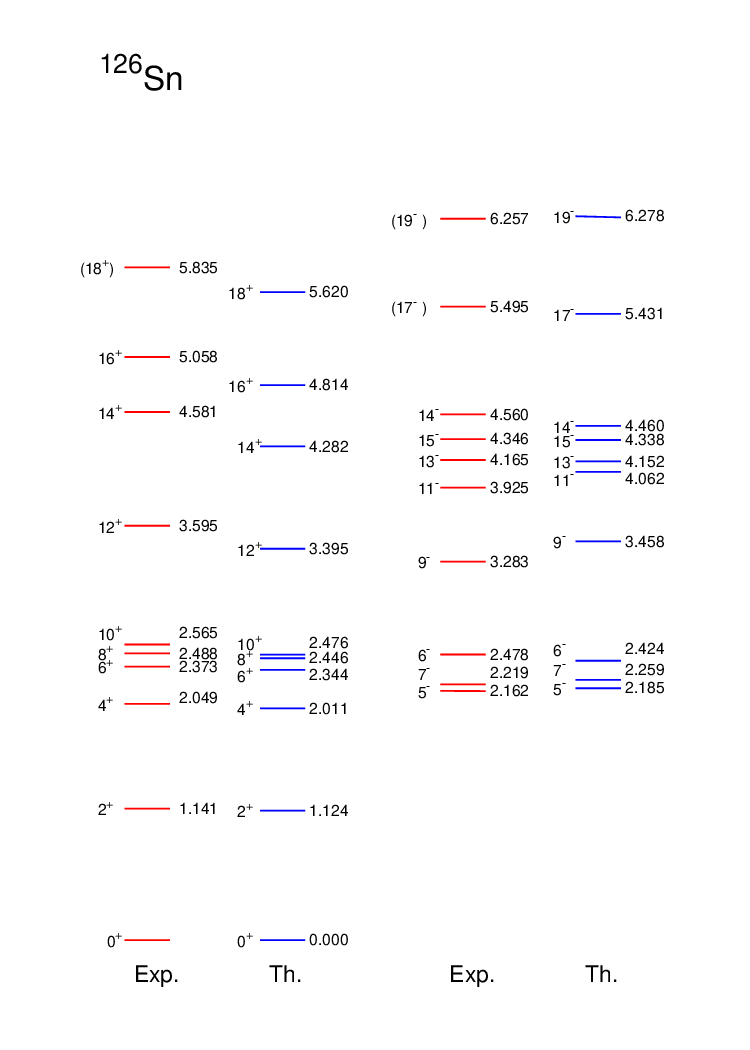}
\caption{
Comparison of experimental \cite{astier,Iskra} and calculated excitation spectra for $^{126}$Sn 
using SN100PN interaction.}
\label{f_sn126}
\end{figure*}
Comparison of the calculated values with the experimental data for $^{126}$Sn is shown in
figure~\ref{f_sn126}.
The shell model calculation predicts
energies of the 2$^+$ and 4$^+$  levels only 17 keV and 38 keV lower, respectively,
than the experimental ones. This shows 
slightly better agreement as compared to that 
for $^{124}$Sn. The values of the gaps between 
4$^+$ and 6$^+$ are 324 and 
333 keV in the experiment and calculation, respectively. They
are also in better agreement with the experiment than in $^{120,122,124}$Sn.

\subsection{\label{odd} Analysis of spectra of odd isotopes of Sn}

For the odd isotopes of Sn unpaired neutron interchanges the position of 
the positive and negative parity bands as compared to the even-even isotopes. 
Calculation gives 11/2$^-$ as the ground state for the 
all odd isotopes of Sn considered here. 
For  $^{121,23}$Sn, in the experiment, the $11/2^-$ level is slightly higher than
the ground states of these nuclei. 


\begin{figure*}
\includegraphics[width=15.0cm]{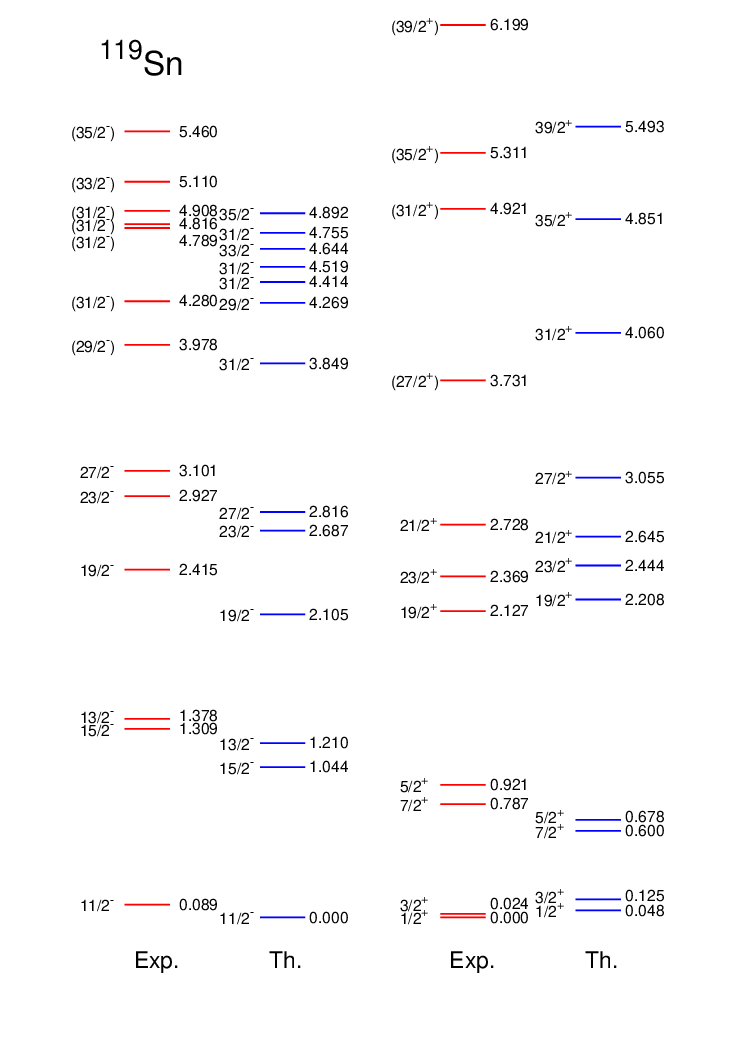}
\caption{
Comparison of experimental \cite{astier,Iskra1} and calculated excitation spectra for $^{119}$Sn 
using SN100PN interaction.}
\label{f_sn119}
\end{figure*}

For the $^{119}$Sn in Fig.\ref{f_sn119} we have presented the calculation up to 35/2$^-$  and 39/2$^+$.
 The calculation gives 11/2$^-$ as the ground state of $^{119}$Sn, 
while in the experiment 11/2$^-$ is the excited state with 89 keV energy. 

\begin{figure*}
\includegraphics[width=15.0cm]{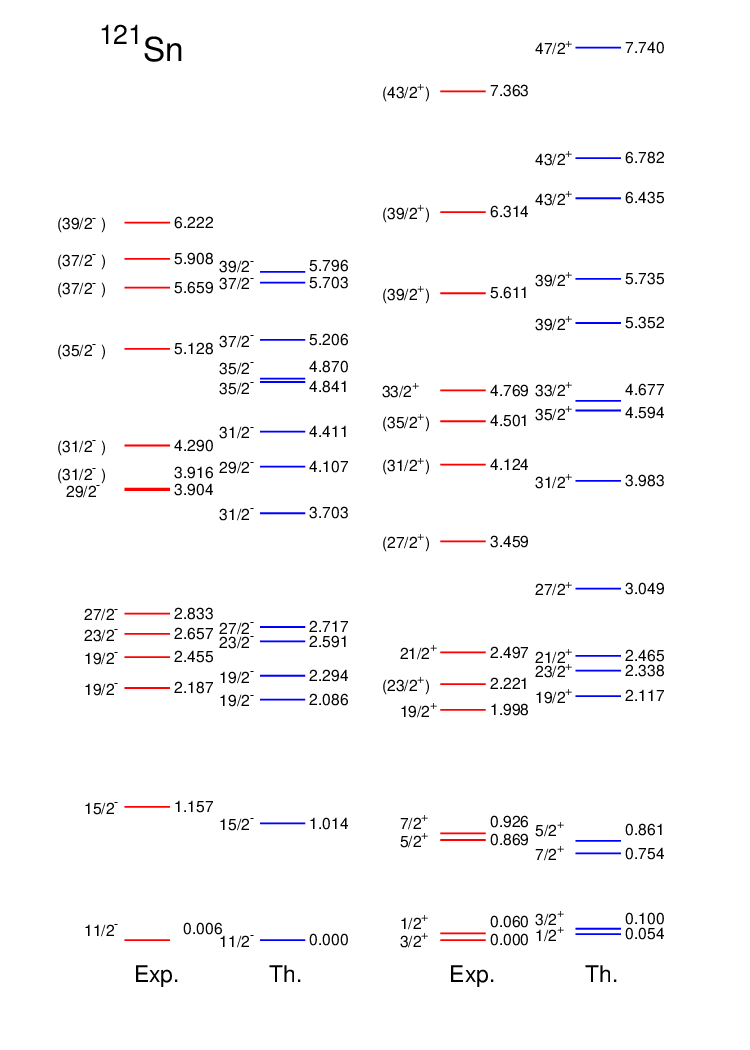}
\caption{
Comparison of experimental \cite{astier,Iskra1} and calculated excitation spectra for $^{121}$Sn 
using SN100PN interaction.}
\label{f_sn121}
\end{figure*}

The calculated values of the $13/2^-$, $15/2^-$, $19/2^-$,
$23/2^-$, $27/2^-$ levels are 168 keV, 265 keV, 310 keV, 240 keV and 285 keV lower than their respective experimental counterparts,
though the calculated pattern of the $^{119}$Sn spectrum is very similar to the experimental
one. 
The spectrum of $^{121}$Sn is given in Fig.~\ref{f_sn121}.
The calculation gives $11/2^-$ as the ground state of $^{121}$Sn, while in the experiment 
the energy of this level is 6 keV. 

The calculated values of the $15/2^-$, $19/2^-_1$, $19/2^-_2$,
$23/2^-$, $27/2^-$, $31/2^-_1$ levels are 143 keV, 
101 keV, 161 keV, 66 keV, 116 keV, 213 keV lower than their experimental counterparts. The calculated $29/2^-$ level is 203 keV higher than that of the experiment. Only the calculated value of this level is larger, otherwise
the calculated pattern of the $^{121}$Sn spectrum is 
now similar to the experimental
one up to $31/2^-_1$ level,
while it was up to $27/2^-$ in the case of $^{119}$Sn. 
More experimental data are available for the positive parity levels of $^{121}$Sn as compared to $^{119}$Sn. 
The calculated pattern is similar to the experimental one.

\begin{figure*}
\includegraphics[width=15.0cm]{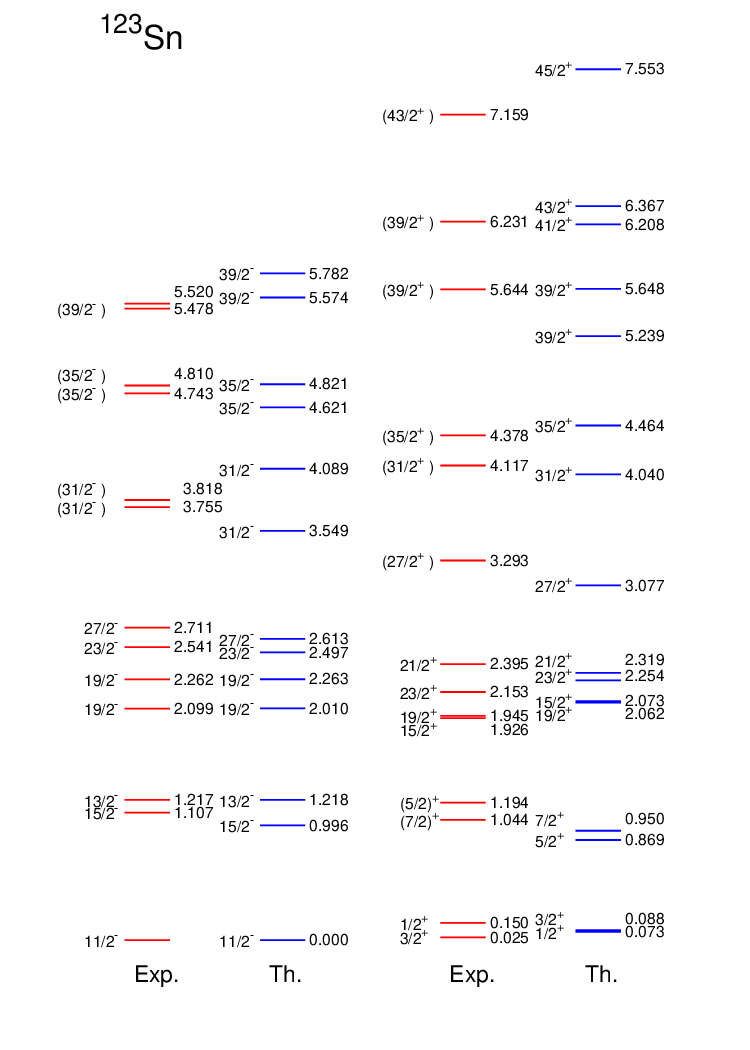}
\caption{
Comparison of experimental \cite{astier,Iskra1} and calculated excitation spectra for $^{123}$Sn 
using SN100PN interaction.}
\label{f_sn123}
\end{figure*}

The spectrum of $^{123}$Sn is given in Fig.~\ref{f_sn123}.
Reaching to $^{123}$Sn isotope one can see that both calculated and experimental
ground state  is $11/2^-$, while for the $^{119,121}$Sn $11/2^-$ 
experimental levels energy values were 89 keV and 6 keV, respectively. 
By careful comparison of the experimental and calculated patterns it can be seen that the whole 
calculated negative parity spectrum is very similar to the experimental one, while the similarity was up to $27/2^-$ and $31/2^-$
for $^{119,121}$Sn, respectively.
The calculated values of $15/2^-_1$, $19/2^-_1$,
$23/2^-$, $27/2^-$, $31/2^-_1$ levels are  111 keV, 
252 keV, 44 keV, 98 keV, 206 keV lower than their experimental counterparts, respectively, while calculated $13/2^-$, $19/2^-_2$, $31/2^-_2$, $35/2^-_2$, $39/2^-_1$, $39/2^-_2$  
levels are 1 keV, 1 keV,  271 keV, 11 keV, 96 keV, 262 keV higher than the experimental ones. If one follows the pattern, $39/2^-$  spin can be assigned to the experimental level at 5520 keV almost definitely according to the shell model prediction. 
From these differences it is also seen that agreement
of the calculated values of the energy levels of $^{123}$Sn are much better than those of $^{119,121}$Sn.
All negative and positive parity levels of $^{123}$Sn are better described as compared to $^{119,121}$Sn by the shell model calculation.

\begin{figure*}
\includegraphics[width=15.0cm]{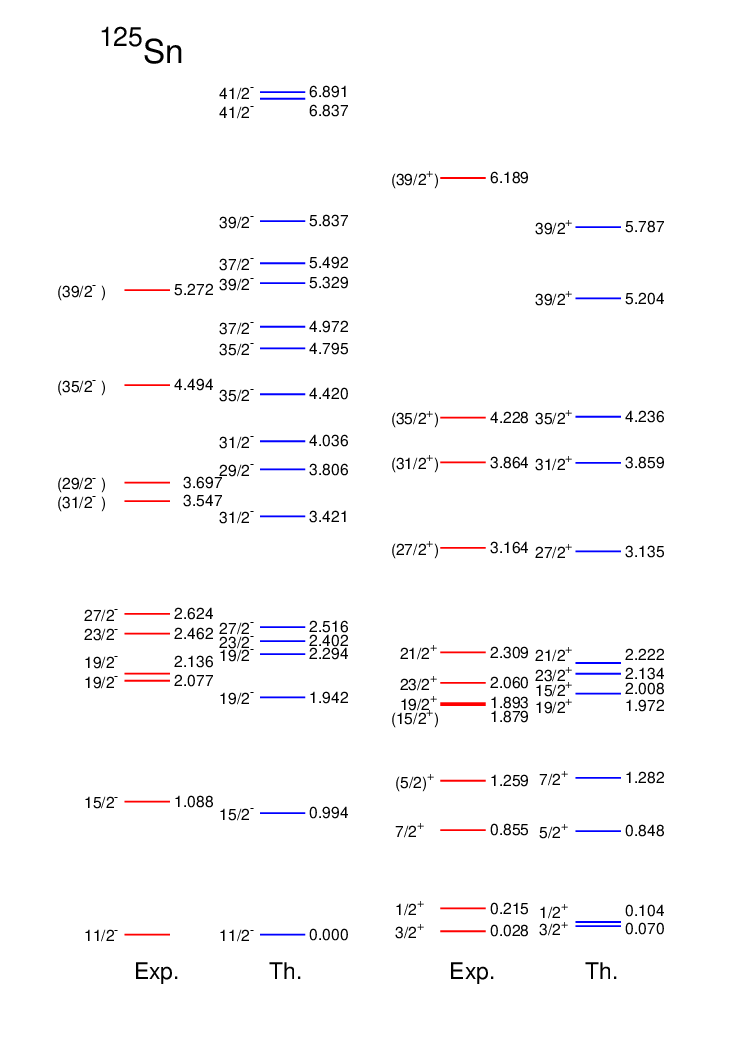}
\caption{
Comparison of experimental \cite{astier,Iskra1} and calculated excitation spectra for $^{125}$Sn 
using SN100PN interaction.}
\label{f_sn125}
\end{figure*}

As is seen from figure ~\ref{f_sn125}, both calculated and experimental
ground states  are $11/2^-$ as it was for $^{123}$Sn. All respective positive and negative parity
excited states energies are lower both in the experiment and calculation with respect to ground state
as compared
to $^{119,121}$Sn isotopes.

Similarity of the experimental and calculated patterns is kept. However, between the experimental $(31/2)^-$ and $(35/2)^-$
levels the second predicted $31/2^-_2$ appears.   
The calculated $39/2^-_1$ level is only 57 keV higher than the experimental one. In the calculation, below this level,
there are the $35/2_2^-$ and $37/2^-_1$ levels and above this level there are the $37/2_2^-$ and $39/2^-_2$ levels. The calculated values of the $15/2^-$, $19/2^-_1$,
$23/2^-$, $27/2^-$, $31/2^-$ are 94 keV, 
135 keV, 60 keV, 108 keV, 126 keV levels lower than their experimental counterparts. The $19/2^-_2$ and $29/2^-$ levels are 158 keV, 
109 keV higher than the experimental one. 
From these differences it also is seen that agreement
of the calculated values of the energy levels of $^{125}$Sn are  better than those of $^{119,121,123}$Sn.

\section{Configuration of the isomeric states}
$^{119-126}$Sn isotopes, which contain more than 68 neutrons, are good for studying high spin states since they do 
contain $\nu (h_{11/2})^n$  with $v=4,5,6$ and the high spin states cannot be formed only by  $\nu s_{1/2}$ and $\nu d_{3/2}$ orbitals themselves.    
In Ref.~\cite{astier}, the $^{119-126}$Sn isotopes have been produced as fragments
of binary fission induced by heavy ions.  Among
them, isomeric states have been established from the delayed
coincidences between fission fragment detectors and the gamma array.
Further experimental data are taken from Refs. \cite{Iskra,Iskra1}.
All the observed states treated in  terms of broken neutron pairs occupying the $\nu (h_{11/2})$ orbital. 
The configurations of high-spin states are due to $(\nu h_{11/2})^n$  or $(\nu h_{11/2})^n$$(\nu d_{3/2})^1$ ones.

Configurations of the isomeric states $10^+$, $5^-$, $7^-$, $15^-$ and $19^-$  of even isotopes of Sn and $27/2^-$, $19/2^+$ and $23/2^+$ 
odd isotopes, emerging from the current shell model calculation, are given separately in Table 1. The seniorities given in this Table are
proposed in~\cite{astier}.   As is seen from Table 1, the percentage of the seniority isomers configurations increases by increasing the neutrons number.
This is because the role of $h_{11/2}$ orbital is dominating and average occupancy of this orbital is increasing.
The average occupancy of $h_{11/2}$ orbital is 5.17 in $^{119}$Sn and 7.79 in $^{125}$Sn.
 The wave functions are very fragmented, for example,  different configurations corresponding to $10^+$ state
 in $^{120}$Sn are shown in the Table 2.
The $10^+$ states of all even isotopes 
and $27/2^-$ states of all odd isotopes are formed by breaking pairs in pure
$\nu (h_{11/2})$ orbital with $v=2$ and $v=3$, respectively. The $d_{3/2}$ and $s_{1/2}$ orbitals also participate in the formation of other isomeric states of the
Sn isotopes. 

In the case of $^{120,122,124,126}$Sn isotopes the different seniority states are coming due to maximally-aligned spin, which involve successive pair breakings in the $\nu (h_{11/2})$ orbit.
For 10$^+$    $(h_{11/2}^2)$, here  one pair breaking of $h_{11/2}$ orbital, thus, seniority is two
($v =2$);  $ 5^-$    $(h_{11/2}^1s_{1/2}^1)$, here  one unpaired neutron is in $h_{11/2}$ orbital and one neutron is in $s_{1/2}$ orbital, thus, seniority is two ($v =2$);  $ 7^-$   $(h_{11/2}^1d_{3/2}^1)$, here  one unpaired neutron is in $h_{11/2}$ orbital and one neutron is in $d_{3/2}$ orbital,
thus, seniority is two ($v =2$);  $ 15^-$   $(h_{11/2}^3d_{3/2}^1)$, here  three unpaired neutrons are in $h_{11/2}$ orbital and one neutron is in $d_{3/2}$ orbital  
the number of unpaired  neutrons are 4, thus, seniority is four ($v =4$).
For $ 19^-$  $(h_{11/2}^5d_{3/2}^1)$, here  five unpaired neutrons are in $h_{11/2}$ orbital and one neutron is in $d_{3/2}$ orbital,
thus, seniority is six ($v = 6$).\\


In the case of $^{119,121,123,125}$Sn isotopes the different seniority states are coming due to maximally-aligned spin, which involve successive pair breakings in the $\nu (h_{11/2})$ orbit.
For $ 27/2^-$   $(h_{11/2}^3)$, here  three unpaired neutrons are in $h_{11/2}$ orbital,  thus,  seniority is three ($v = 3)$;  $ 19/2^+$   $(h_{11/2}^2s_{1/2}^1)$, here  two unpaired neutrons are in $h_{11/2}$ orbital and one neutron is in $s_{1/2}$ orbital, thus,  seniority is three ($v = 3)$;  $ 23/2^+$   $(h_{11/2}^2d_{3/2}^1)$, here  two unpaired neutrons are in $h_{11/2}$ orbital and one neutron is in $d_{3/2}$ orbital,
thus,  seniority is three ($v = 3)$;   $ 35/2^+$   $(h_{11/2}^4d_{3/2}^1)$, here  four unpaired neutrons are in $h_{11/2}$ orbital and one neutron is in $d_{3/2}$ orbital, thus,  seniority is five ($v = 5)$.
The seniority of $39/2^+$ state is seven ($v=7$). Here we have six unpaired neutrons are in $h_{11/2}$ orbital and one neutron is in $d_{3/2}$ orbital,  the number of unpaired  neutrons are 7, thus, seniority is seven ($v = 7)$.  


\section{Transition probabilities}
\vspace{0.2cm}
In the Table~3, we have shown $B(E2)$ transition probabilities of the isomeric states for $^{119-126}$Sn isotopes. 
These  calculated values of $E2$ transition probabilities are important for future experiments. 
After calculating $B(E2; 0^+ \rightarrow 2^+)$  for heavier Sn isotopes, we have found that the calculated results
showing good agreement with experimental data with $e_n=0.8e$ (see Fig. 11).
Thus for comparison apart from standard effective charge for neutron $e_n=0.5e$, we have also calculated $B(E2)$ values with
second set of effective charge $e_n=0.8e$, although for high-spin states $B(E2)$ values  are very large with this effective charge.
The SN100PN interaction results for energy levels are in a reasonable agreement with the experimental data, however the 
calculated results for $B(E2)$ transitions are not in good agreement. 
This  might be due to a deficiency in the wave functions rather than
in the effective operator. Because electromagnetic properties are sensitive to the detailed composition of the nuclear wave function.

\begin{figure*}
\vspace{+10cm}
\begin{center}
\includegraphics[width=13.0cm]{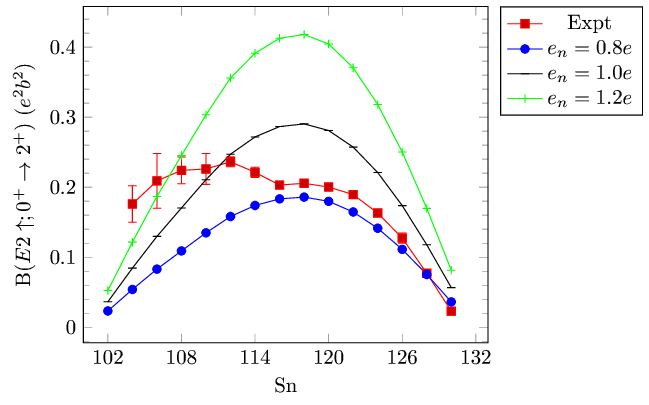}
\caption{
Shell model results  of $B(E2)$ values for Sn isotopes.}
\end{center}
\end{figure*}

\begin{figure*}
\vspace{+10cm}
\begin{center}
\includegraphics[width=9.0cm]{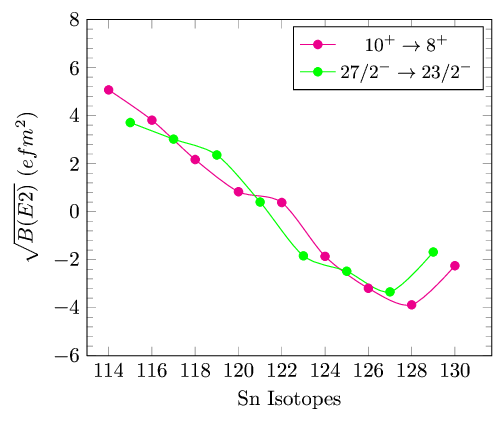}
\caption{
Shell model results  (with $e_n=0.5e$) for amplitudes of the reduced transition probabilities for
$E2$ isomeric decay in Sn isotopes. The ${B(E2;27/2^- \rightarrow  23/2^-)}$  values
are multiplied by the factor 0.264.}
\end{center}
\end{figure*}

In the Fig. 12, we have shown amplitudes of the reduced transition probabilities for
$E2$ isomeric decay in Sn isotopes corresponding to isomeric states $10^+$ and $27/2^-$. 
The $10^+$ and $8^+$ states are due to seniority two and $27/2^-$ and $23/2^-$ states are due to seniority three.
Thus  linear decrease in $B(E2)$ values are obtained. The behaviour of $\sqrt{B(E2;10^+ \rightarrow  8^+)}$  and
$\sqrt{B(E2;27/2^- \rightarrow  23/2^-)}$ 
transitions as a
function of mass number (A) was used to determine the filling of the $\nu h_{11/2}$ orbital. 
Since we have plotted results in the same figure thus we have
multiplied by factor 0.264 for $B(E2)$ values corresponding to  $27/2^- \rightarrow  23/2^-$ transition, this is coming from expressions for the $v=2$ and $v=3$ transitions as reported in
Ref. \cite{PRLMcNeill} (for more details about the expressions please see this reference).
The 0.264 factor is used to compensate coefficients of fractional parentage (c.f.p.) entering in the expression \cite{plb} of states with seniority 
$v =3$ (for $23/2^-$ and $27/2^-$) in comparison to states with seniority 
$v =2$ (for $8^+$ and $10^+$). 
The similar behavior of $B(E2)$ values for $10^+ \rightarrow  8^+$ and $27/2^- \rightarrow  23/2^-$ transitions
is coming because initial and final 
states are having same seniority with increasing neutron numbers.
\section{Summary and Conclusion}
\vspace{-0.5cm}
 In the present work we have reported comprehensive shell model description of high-spin states in $^{119-126}$Sn isotopes. 
With the shell model configurations of different high-spin states, we have analyzed different isomeric states in these nuclei.
These states can be described in terms of several broken neutron pairs occupying the $h_{11/2}$ orbital.
The high-spin isomers in Sn isotopes are due to seniority ($v$) = 2, 3, 4, and 5. 
For the  $^{120,122,124,126}$Sn isotopes, the seniority of isomeric states $10^+$, $5^-$ and $7^-$   are two ($v$ = 2);
seniority of isomeric state $15^-$ is four ($v$ = 4); the seniority of $19^-$ state is six ($v$ = 6).
For the  $^{119,121,123,125}$Sn isotopes, the seniority of isomeric states $19/2^+$, $23/2^+$, and $27/2^-$  are three ($v$ = 3);
the seniority of isomeric state $35/2^+$ is five ($v=5$); the seniority of $39/2^+$ state is seven ($v=7$).
Although, the probability of different configurations are not so large but they are increasing with neutron numbers for heavier Sn isotopes.
We have also reported $B(E2)$ values for different transitions for $^{120-126}$Sn isotopes.

{\bf Acknowledgement}\\
B. Bhoy acknowledges financial support from MHRD (Govt. of India) for her Ph.D. thesis work.
P.C.S. acknowledges the hospitality extended to him during his stay at the Saitama University, Japan. 
MJE's work is partially supported by the Uzbekistan National Agency of Science and Technology (OT-F2-15).
We would like to thank N. Yoshinaga and  K. Yanase for useful discussions and initial collaboration in this work.

\begin{landscape}
\begin{table}[hbtp]
\caption{  Configurations of isomeric states in $^{119,120,121,122,123,124,125,126}$Sn isotopes. Also the probability of largest component of the configuration are given in the bracket.}
\label{t_be2}
\begin{center}
\begin{tabular}{rrrccc}
\hline       
\hline
 Spin    &   Seniority & \hspace{0.5cm} $^{120}$Sn    &  $^{122}$Sn    &$^{124}$Sn     & $^{126}$Sn \\
\hline 
$ 10^+$ & $  v =2$ $(h_{11/2}^2)$ & $g_{7/2}^6d_{5/2}^6d_{3/2}^2h_{11/2}^{6}$ [10.13\%]  &  $g_{7/2}^8d_{5/2}^6d_{3/2}^2h_{11/2}^{6}$ [15.07\%]  & $g_{7/2}^8d_{5/2}^6d_{3/2}^2s_{1/2}^2h_{11/2}^{6}$ [18.64\%]   & $g_{7/2}^8d_{5/2}^6d_{3/2}^2s_{1/2}^2h_{11/2}^{8}$ [37.24\%] \\ 

$ 5^- $ & $ v =2$ $(h_{11/2}^1s_{1/2}^1)$  &   $g_{7/2}^8d_{5/2}^6s_{1/2}^1h_{11/2}^{5}$ [11.79\%]   &  $g_{7/2}^8d_{5/2}^6d_{3/2}^2s_{1/2}^1h_{11/2}^{5}$ [12.26\%] &  $g_{7/2}^8d_{5/2}^6d_{3/2}^2s_{1/2}^1h_{11/2}^{7}$ [23.07\%]   &  $g_{7/2}^8d_{5/2}^6d_{3/2}^2s_{1/2}^1h_{11/2}^{9}$ [27.67\%] \\ 

$ 7^- $ & $ v =2$ $(h_{11/2}^1d_{3/2}^1)$  & $g_{7/2}^8d_{5/2}^6d_{3/2}^{1}h_{11/2}^{5}$ [17.00\%]   &   $g_{7/2}^8d_{5/2}^6d_{3/2}^{1}h_{11/2}^{7}$ [16.24\%] & $g_{7/2}^8d_{5/2}^6d_{3/2}^{1}s_{1/2}^2h_{11/2}^{7}$ [22.14\%] & $g_{7/2}^8d_{5/2}^6d_{3/2}^{3}s_{1/2}^2h_{11/2}^{7}$ [24.63\%] \\ 

 $15^-$   & $ v =4$ $(h_{11/2}^3d_{3/2}^1)$     &  $g_{7/2}^8d_{5/2}^6d_{3/2}^{1}h_{11/2}^{5}$ [27.60\%]   & $g_{7/2}^8d_{5/2}^6d_{3/2}^{1}h_{11/2}^{7}$ [23.21\%]  & $g_{7/2}^8d_{5/2}^6d_{3/2}^{1}s_{1/2}^2h_{11/2}^{7}$ [28.76\%] & $g_{7/2}^8d_{5/2}^6d_{3/2}^{3}s_{1/2}^2h_{11/2}^{7}$[30.86\%]\\


\hline
\hline

 Spin    &   Seniority    &\hspace{0.5cm} $^{119}$Sn   &  $^{121}$Sn     &$^{123}$Sn    & $^{125}$Sn \\
\hline 
$27/2^-$ & $ v = 3$ $(h_{11/2}^3)$ & $g_{7/2}^8d_{5/2}^6h_{11/2}^{5}$  [13.80\%]   &  $g_{7/2}^8d_{5/2}^6d_{3/2}^2h_{11/2}^{5}$  [14.26\%]  & $g_{7/2}^8d_{5/2}^6d_{3/2}^2h_{11/2}^{7}$  [19.86\%]   & $g_{7/2}^8d_{5/2}^6d_{3/2}^2s_{1/2}^2h_{11/2}^{7}$  [34.55\%] \\ 

$19/2^+ $ & $ v = 3$ $(h_{11/2}^2s_{1/2}^1)$ &  $g_{7/2}^8d_{5/2}^6s_{1/2}^1h_{11/2}^{4}$  [11.77\%]  &  $g_{7/2}^8d_{5/2}^6s_{1/2}^1h_{11/2}^{6}$  [13.81\%]   & $g_{7/2}^8d_{5/2}^6d_{3/2}^2s_{1/2}^1h_{11/2}^{6}$  [17.74\%]   & $g_{7/2}^8d_{5/2}^6d_{3/2}^2s_{1/2}^1h_{11/2}^{8}$  [31.15\%] \\ 

$ 23/2^+$ & $ v = 3$ $(h_{11/2}^2d_{3/2}^1)$  & $g_{7/2}^8d_{5/2}^6d_{3/2}^1h_{11/2}^{4}$  [18.49\%]  &  $g_{7/2}^8d_{5/2}^6d_{3/2}^1h_{11/2}^{6}$  [21.77\%]  & $g_{7/2}^8d_{5/2}^6d_{3/2}^1s_{1/2}^2h_{11/2}^{6}$  [20.19\%]  & $g_{7/2}^8d_{5/2}^6d_{3/2}^1s_{1/2}^2h_{11/2}^{8}$  [25.57\%]  \\ 
  
$ 35/2^+$ & $ v = 5$ $(h_{11/2}^4d_{3/2}^1)$  &  &   & $g_{7/2}^8d_{5/2}^6d_{3/2}^3s_{1/2}^2h_{11/2}^{4}$  [2.50\%]  &  \\ 
\hline
\end{tabular}
\end{center}
\end{table}
\end{landscape}
\begin{table}
\caption{ Probability of different configurations for $10^+$ state in $^{120}$Sn. }
\begin{tabular}{rr}
\hline       
\hline         
    Probability &  Wave functions \\
    \hline
    8.4$\%$   &  $\nu(g_{7/2}^8d_{5/2}^6d_{3/2}^2h_{11/2}^4)$ \\
    5.9$\%$   &   $\nu(g_{7/2}^8d_{5/2}^6s_{1/2}^2h_{11/2}^4)$ \\
    9.6$\%$   &   $\nu(g_{7/2}^8d_{5/2}^6h_{11/2}^6)$ \\
    3.1$\%$   &   $\nu(g_{7/2}^8d_{5/2}^4d_{3/2}^2$ $s_{1/2}^2h_{11/2}^4)$ \\
    6.6$\%$   &  $\nu(g_{7/2}^8d_{5/2}^4d_{3/2}^2h_{11/2}^6)$ \\
    4.0$\%$   &  $\nu(g_{7/2}^8d_{5/2}^4s_{1/2}^2h_{11/2}^6)$ \\
    2.4$\%$   &  $\nu(g_{7/2}^8d_{5/2}^4h_{11/2}^8)$ \\
    1.9$\%$   &  $\nu(g_{7/2}^8d_{5/2}^2d_{3/2}^2s_{1/2}^2h_{11/2}^6)$ \\
    1.8$\%$   &  $\nu(g_{7/2}^6d_{5/2}^6d_{3/2}^4h_{11/2}^4)$ \\
    4.7$\%$   &  $\nu(g_{7/2}^6d_{5/2}^6d_{3/2}^2s_{1/2}^2h_{11/2}^4)$ \\
    10.1$\%$  &  $\nu(g_{7/2}^6d_{5/2}^6d_{3/2}^2h_{11/2}^6)$ \\
    4.9$\%$   &  $\nu(g_{7/2}^6d_{5/2}^6s_{1/2}^2h_{11/2}^6)$ \\
    2.6$\%$   &  $\nu(g_{7/2}^6d_{5/2}^6h_{11/2}^8)$ \\
    1.8$\%$   &  $\nu(g_{7/2}^6d_{5/2}^4d_{3/2}^4h_{11/2}^6)$ \\
    4.5$\%$   &  $\nu(g_{7/2}^6d_{5/2}^4d_{3/2}^2s_{1/2}^2h_{11/2}^6)$ \\
    2.9$\%$   &  $\nu(g_{7/2}^6d_{5/2}^4d_{3/2}^2h_{11/2}^8)$ \\
    1.4$\%$   &  $\nu(g_{7/2}^6d_{5/2}^4s_{1/2}^2h_{11/2}^8)$ \\
    1.3$\%$   &  $\nu(g_{7/2}^4d_{5/2}^6d_{3/2}^2s_{1/2}^2h_{11/2}^6)$ \\
        \hline
    \hline
    \end{tabular}
\end{table}

 \begin{landscape}   
\begin{table}
\caption{  Calculated $B(E2)$ values for different transitions in $e^2fm^4$ using $e_n=0.5e$ ; $e_n=0.8e$  separated by ``/".
Experimental data have been taken from Ref. \cite{Iskra}.}
\label{t_be2}
\begin{center}
\begin{tabular}{rrccccccc}
\hline       
\hline
                     & $^{120}$Sn &\hspace{1.0cm}~~~~~    &\hspace{1.0cm}~~~~~ $^{122}$Sn  &  &\hspace{1.0cm}~~~~~ $^{124}$Sn  & &~~~~~  $^{126}$Sn &\\
${B(E2; i \rightarrow   f}$)  & EXPT. & SM   & EXPT. & SM    & EXPT. & SM    & EXPT. & SM  \\
\hline 
$10^+ \rightarrow  8^+$  & NA      & 0.69/1.50     &  NA       & 0.15/0.47      &  NA        &  3.46/8.75     &   NA    & 10.23/26.38 \\ 
$15^- \rightarrow  13^-$ & 21(2) & 81.11/211.89  & 4.7(4)  & 76.83/192.00   &  3.2(4)  &  34.97/88.17   & 22(2) & 14.22/36.38 \\ 
$7^- \rightarrow  5^-$   & NA      & 28.23/74.92   &  NA       & 16.42/39.66    &   NA       &  4.94/11.59    &  NA     & 0.0013/0.003 \\ 
$5^- \rightarrow  3^-$   &  NA     & 39.73/102.74  &  NA       & 45.86/118.79   &  NA        & 44.67/113.31   &  NA     & 21.18/53.74\\ 

\hline       
\hline   
                     & $^{119}$Sn &    &$^{121}$Sn  &  &$^{123}$Sn  & & $^{125}$Sn &\\
${B(E2; i \rightarrow   f}$)  & EXPT. & SM   & EXPT. & SM    & EXPT. & SM    & EXPT. & SM  \\
\hline 
$27/2^- \rightarrow   23/2^-$ & 73(7) & 21.09/57.52   & NA     & 0.61/1.62  & NA      &  12.86/32.81  & NA & 23.36/60.42 \\ 
$23/2^+ \rightarrow   19/2^+$ & 6.7(8) & 59.12/149.82  & 1.8(2) & 52.17/133.33 & 0.22(3) &  25.31/43.15  & 5.4(7) & 6.95/21.99 \\ 
$19/2^+ \rightarrow   15/2^+$ & NA     & 0.09/0.91   & NA    & 5.12/19.81   &  6.7(25) &  9.17/24.17  & 22(4) & 10.03/24.60 \\  
$35/2^+ \rightarrow   31/2^+$ & NA     &  10.58/27.09     & NA    &  58.41/149.52    &  109(29) &   0.391/1.003  & NA  &  47.67/122.04 \\ 
\hline
\end{tabular}
\end{center}
\end{table}
\end{landscape}

\bibliographystyle{npa}

\begin{thebibliography}{10}
\providecommand{\url}[1]{\texttt{#1}}
\providecommand{\urlprefix}{URL }
\providecommand{\eprint}[2][]{\url{#2}}


\bibitem{astier}
A.Astier, M.-G.Porquet, Ch.Theisen, D.Verney, I.Deloncle, M.Houry, R.Lucas, F.Azaiez, G.Barreau, D.Curien, O.Dorvaux, G.Duchene,
B.J.P.Gall, N.Redon, M.Rousseau, and O.Stezowski, {\color{blue} Phys. Rev. C \textbf{85},  054316 (2012)}.

\bibitem{Iskra}  \L{}.~W. Iskra, R.Broda, R.V.F.Janssens, J.Wrzesinski, B.Szpak, C.J.Chiara, M.P.Carpenter, B.Fornal, N.Hoteling, F.G.Kondev, W.Krolas, T.Lauritsen, 
T.Pawlat, D.Seweryniak, I.Stefanescu, W.B.Walters, and S.Zhu,  {\color{blue} Phys. Rev. C \textbf{89},  044324 (2014)}.  
  
 \bibitem{Iskra1}  \L{}.~W. Iskra, L.W.Iskra, R.Broda, R.V.F.Janssens, C.J.Chiara, M.P.Carpenter, 
 B.Fornal, N.Hoteling, F.G.Kondev, W.Krolas, T.Lauritsen, T.Pawlat, D.Seweryniak, I.Stefanescu, W.B.Walters, J.Wrzesinski, and S.Zhu,  
 {\color{blue} Phys. Rev. C \textbf{93}, 
  014303 (2016)}. 
   
  
\bibitem{sandlescu} N. Sandulescu, J. Blomqvist, T. Engeland, M. Hjorth-Jensen, A. Holt, R. J. Liotta, and E. Osnes, 
{\color{blue} Phys. Rev. C \textbf{55}, 2708 (1997)}.

\bibitem{qi} C. Qi and Z. X. Xu, 
{\color{blue} Phys. Rev. C \textbf{86}, 044323 (2012)}.

\bibitem{corragio} L. Coraggio, A. Covello, A. Gargano, N. Itaco, and T. T. S. Kuo, 
{\color{blue} Phys. Rev. C \textbf{91}, 041301(R) (2015)}.

\bibitem{togashi} T. Togashi, Y. Tsunoda,  T. Otsuka, N. Shimizu, and M. Honma, 
{\color{blue} Phys. Rev. Lett. \textbf{121}, 062501 (2018)}.


\bibitem{snptep}   H. Wang et al., 
{\color{blue} Prog. Theo. Exp. Phys. \textbf{2014}, 023D02 (2014)}.

\bibitem{biswas}   S. Biswas et al., 
{\color{blue} Phys. Rev. C \textbf{99}, 064302 (2019)}.
  
\bibitem{hinke} C.B.Hinke, M.Bohmer, P.Boutachkov, T.Faestermann, H.Geissel, J.Gerl, R.Gernhauser, M.Gorska, A.Gottardo, H.Grawe, J.L.Grebosz, R.Krucken, N.Kurz, Z.Liu, L.Maier, F.Nowacki, S.Pietri, Zs.Podolyak, K.Sieja, K.Steiger, K.Straub, H.Weick, H.-J.Wollersheim, P.J.Woods, N.Al-Dahan, N.Alkhomashi, A.Atac, A.Blazhev, N.F.Braun, I.T.Celikovic, T.Davinson, I.Dillmann, C.Domingo-Pardo, P.C.Doornenbal, G.de France, G.F.Farrelly, F.Farinon, N.Goel, T.C.Habermann, R.Hoischen, R.Janik, M.Karny, A.Kaskas, I.M.Kojouharov, Th.Kroll, Y.Litvinov, S.Myalski, F.Nebel, S.Nishimura, 
C.Nociforo, J.Nyberg, A.R.Parikh, A.Prochazka, P.H.Regan, C.Rigollet, H.Schaffner, C.Scheidenberger, S.Schwertel, P.-A.Soderstrom, 
S.J.Steer, A.Stolz, and P.Strmen,
 {\color{blue} Nature \textbf{486},  341 (2012)}.

\bibitem{ekstrom}A.Ekstrom, J.Cederkall, C.Fahlander, M.Hjorth-Jensen, F.Ames, P.A.Butler, T.Davinson, J.Eberth, F.Fincke, A.Gorgen, M.Gorska, D.Habs, A.M.Hurst, M.Huyse, O.Ivanov, J.Iwanicki, O.Kester, U.Koster, B.A.Marsh, J.Mierzejewski, P.Reiter, H.Scheit, D.Schwalm, S.Siem, 
G.Sletten, I.Stefanescu, G.M.Tveten, J.Van de Walle, P.Van Duppen, D.Voulot, N.Warr, D.Weisshaar, F.Wenander, and M.Zielinska,  {\color{blue} 
Phys. Rev. Lett. \textbf{101},  012502 (2008)}.

\bibitem{ruiz} R. F. Garcia Ruiz et. al, IS613: Laser Spectroscopy of neutron-deficient Sn isotopes, Proposal INTC-P-456 CERN-INTC-2016-006 (CERN, 2016).

\bibitem{zamick} S.Robinson and  L.Zamick,  {\color{blue} Int.J.Mod.Phys. E \textbf{26}, 1750053 (2017)}.

\bibitem{Morris} T. D. Morris, J. Simonis, S. R. Stroberg, C. Stumpf, G. Hagen, J. D. Holt, G. R. Jansen, T. Papenbrock, R. Roth, and A. Schwenk, 
 {\color{blue} Phys. Rev. Lett. \textbf{120},  152503 (2018)}.
 
 \bibitem{Darb} I. G. Darby, R. K. Grzywacz, J. C. Batchelder, C. R. Bingham, L. Cartegni, C. J. Gross, M. Hjorth-Jensen, D. T. Joss, S. N. Liddick, 
 W. Nazarewicz, S. Padgett, R. D. Page, T. Papenbrock, M. M. Rajabali, J. Rotureau, and K. P. Rykaczewski,  
 {\color{blue} Phys. Rev. Lett. \textbf{105}, 162502 (2010)}.
 
 
\bibitem{talmi} I. Talmi, Simple Models of Complex Nuclei (Harwood, Chur, Switzerland 1993).


\bibitem{piet} P. Van Isacker, {\color{blue} Journal of Physics: Conference Series \textbf{322}, 012003 (2011). }

\bibitem{fant} B. Fant et al., {\color{blue}  Nucl. Phys. A  \textbf{475}, 338 (1987)}.

\bibitem{fotiades} N.Fotiades, M.Devlin, R.O.Nelson, J.A.Cizewski, R.Krucken, R.M.Clark, P.Fallon, I.Y.Lee, A.O.Macchiavelli, and W.Younes,
 {\color{blue} Phys. Rev. C \textbf{84}, 054310 (2011)}.


\bibitem{Mac} R. Machleidt, F. Sammarruca, Y. Song,  {\color{blue} Phys. Rev. C \textbf{53}, R1483 (1996)}.
 
  
\bibitem{PhysRevC.71.044317}
B. A. Brown, N. J. Stone, J. R. Stone, I. S. Towner, and M. Hjorth-Jensen,  {\color{blue} Phys. Rev. C \textbf{71},
 044317  (2005)}.  
  
\bibitem{Antoine}
E. Caurier,  G. Mart\'inez-Pinedo , F. Nowacki, A. Poves, and A. P. Zuker, 
 {\color{blue} Rev.\ Mod.\ Phys. {\bf77}, 427  (2005)}.   

\bibitem{pc1}P.~C. Srivastava, M.J. Ermamatov, I.~O. Morales,  {\color{blue} J. Phys. G \textbf{40},  035106 (2013)}.

\bibitem{pc2}V. Kumar, P.~C. Srivastava, M.~J. Ermamatov, I.~O. Morales,  {\color{blue} Nucl. Phys. A \textbf{942}, 1 (2015)}.

\bibitem{pc3} S.Biswas, R.Palit, J.Sethi, S.Saha, A.Raghav, U.Garg, Md.S.R.Laskar, F.S.Babra, Z.Naik, S.Sharma, A.Y.Deo, V.V.Parkar, B.S.Naidu, R.Donthi, S.Jadhav,
H.C.Jain, P.K.Joshi, S.Sihotra, S.Kumar, D.Mehta, G.Mukherjee, A.Goswami, and P.C.Srivastava,  {\color{blue} Phys. Rev. C \textbf{95},  064320  (2017)}.

 \bibitem{pc4}S. Biswas, R.Palit, A.Navin, M.Rejmund, A.Bisoi, M.S.Sarkar, S.Sarkar, S.Bhattacharyya, D.C.Biswas, M.Caamano, M.P.Carpenter, D.Choudhury, E.Clement, L.S.Danu, O.Delaune, F.Farget, G.de France, S.S.Hota, B.Jacquot,
 A.Lemasson, S.Mukhopadhyay, V.Nanal, R.G.Pillay, S.Saha, J.Sethi, P.Singh, P.C.Srivastava, and S.K.Tandel, 
  {\color{blue} Phys. Rev. C \textbf{93}, 034324  (2016)}.
 
\bibitem{pc5} A. Astier, M.-G. Porquet, G. Duchene, F. Azaiez, D. Curien, I. Deloncle, O. Dorvaux, B.J.P.Gall, M.Houry, R.Lucas, P.C. Srivastava,
N.Redon, M.Rousseau, O.Stezowski, and Ch.Theisen,  {\color{blue} Phys. Rev. C \textbf{87},  054316 (2013) }. 

\bibitem{pc6} J.~D. Vergados, F.~T. Avignone, III, M. Kortelainen, P. Pirinen, P.~C. Srivastava, J. Suhonen, and A.~W. Thomas, 
 {\color{blue} J. Phys. G \textbf{43},  115002 (2016)}.

\bibitem{pc7}P. Pirinen, P.~C. Srivastava, J. Suhonen, and M. Kortelainen,  {\color{blue} Phys. Rev. D \textbf{93},  095012 (2016) }. 

\bibitem{pc8} K. Wimmer, U. Koster, P. Hoff, Th. Kroll, R. Krucken, R. Lutter, H. Mach, Th. Morgan, S. Sarkar,
M. Saha-Sarkar, W. Schwerdtfeger, P.~C. Srivastava, P.~G. Thirolf, and P. Van Isacker,  {\color{blue} Phys. Rev. C \textbf{84},  014329 (2011)}.


\bibitem{PRLMcNeill}
J. H. McNeill,  J. Blomqvist, A.A. Chishti, P.J. Daly, W. Gelletly, M.A.C. Hotchkis, M.  Piiparinen, B.J. Varley, and P.J.  Woods, 
{\color{blue} Phys. Rev. Lett. \textbf{63}, 860 (1989)}.

\bibitem{plb} R.H. Mayer,  D.T. Nisius, I.G. Bearden, P. Bhattacharyya, L. Richter, M. Sferrazza, Z.W. Grabowski, P.J. Daly, R. 
Broda, B. Fornal, I. Ahmad, M.P. Carpenter, R.G. Henry,   R.V.F. Janssens, T.L. Khoo, T. Lauritsen,  Y. Liang and J. 
Blomqvist,  {\color{blue} Phys. Lett. B  \textbf{336}, 308 (1994)}.






\end{thebibliography}

\end{document}